\pdfoutput = 1

\documentclass[%
 reprint,
superscriptaddress,
twocolumn,
 amsmath, amssymb,
 aps,
 prl,
]{revtex4-1}

\usepackage{graphicx}% Include figure files
\usepackage{dcolumn}% Align table columns on decimal point
\usepackage{bm}% bold math
\usepackage{float}
\usepackage[colorlinks = true,
            linkcolor = blue,
            urlcolor  = blue,
            citecolor = blue,
            anchorcolor = black]{hyperref}% add hypertext capabilities
\usepackage{pgffor}

\begin{document}

\preprint{APS/123-QED}

\title{Chaos and correlated avalanches\\in excitatory neural networks with synaptic plasticity}

\author{Fabrizio Pittorino}
\affiliation{Dipartimento di Scienze Matematiche, Fisiche e Informatiche, Universit\`a di Parma, via G.P. Usberti, 7/A - 43124, Parma, Italy}
\affiliation{INFN, Gruppo Collegato di Parma, via G.P. Usberti, 7/A - 43124, Parma, Italy}
\author{Miguel Ib\'a\~nez-Berganza}
\affiliation{INFN, Gruppo Collegato di Parma, via G.P. Usberti, 7/A - 43124, Parma, Italy}
\author{Matteo di Volo}
\affiliation{Group for Neural Theory, Laboratoire de Neurosciences Cognitives, INSERM U960, {\'E}cole Normale Sup\'erieure, Paris, France}
\author{Alessandro Vezzani}
\affiliation{IMEM-CNR, Parco Area delle Scienze, 37/A-43124 Parma, Italy}
\affiliation{Dipartimento di Scienze Matematiche, Fisiche e Informatiche, Universit\`a di Parma, via G.P. Usberti, 7/A - 43124, Parma, Italy}
\author{Raffaella Burioni}
\affiliation{Dipartimento di Scienze Matematiche, Fisiche e Informatiche, Universit\`a di Parma, via G.P. Usberti, 7/A - 43124, Parma, Italy}
\affiliation{INFN, Gruppo Collegato di Parma, via G.P. Usberti, 7/A - 43124, Parma, Italy}

\begin{abstract}
A collective chaotic phase with power law scaling of activity events is observed in a disordered mean field network of purely excitatory leaky integrate-and-fire neurons with short-term synaptic plasticity.  The dynamical phase diagram exhibits two transitions from quasi-synchronous and asynchronous regimes  to the nontrivial, collective, bursty  regime with avalanches. In the homogeneous case without disorder, the system synchronizes and the bursty behavior is reflected into a period doubling transition to chaos for a two dimensional discrete map. Numerical simulations show that the bursty chaotic phase with avalanches exhibits a spontaneous emergence of persistent time correlations and enhanced Kolmogorov complexity. Our analysis reveals a mechanism for the generation 
of irregular avalanches that emerges from the combination of disorder and deterministic underlying chaotic  dynamics.                 
\end{abstract}

%\pacs{Valid PACS appear here}% PACS, the Physics and Astronomy Classification Scheme.
%\keywords{Suggested keywords}%Use showkeys class option if keyword display desired

\maketitle

Networks of spiking neurons feature a wide range of dynamical collective behaviors, that are believed to be crucial for brain functioning \cite{doi:10.1146/annurev.neuro.28.061604.135637}.  Next to uncorrelated and asynchronous dynamics, quasi-synchronous phases and regimes of irregular activity have been observed, showing a still unexplained degree of correlation that could encode part of the neural function \cite{hulata2005self,shu2003turning,wang2010neurophysiological,destexhe2007corticothalamic,10.3389/fnsys.2014.00166,sanchez2000cellular}.  Understanding the mechanisms that generate such experimentally observed collective behaviors and the transition between them is a major goal 
in theoretical neuroscience \cite{doi:10.1146/annurev.neuro.28.061604.135637,PhysRevE.48.1483,PhysRevE.54.5522,PhysRevX.5.041030,Brunel2000,kirst2009sequential,montbrio2015macroscopic,hansel2003asynchronous,panzeri2001role}. 
A  particularly interesting dynamical signature of collective irregular regimes are {\it avalanches} or bursts of spiking neurons with heavy-tailed 
 distributions  of activity \cite{10.3389/fnsys.2014.00166,mejias2010irregular,Livi201360}. Interestingly, in cortical networks, irregular activity at the collective level  \cite{Burns211, London2010} and avalanches characterized by power law distributions 
have been widely observed both $\emph{in\,vitro}$ and $\emph{in\,vivo}$ \cite{Beggs03122003,segev2001observations,petermann2009spontaneous,el2009network}. These regimes are thought to be closely related to information processing in 
the cortex \cite{kinouchi, Ostojic2014, PhysRevX.2.041007} 
and to adaptive \cite{Chialvo2010} and healthy \cite{10.3389/fnsys.2015.00022} behavior. 

Several mechanisms leading to irregular dynamics  and bursts in networks of spiking neurons have been proposed. Irregular dynamical phases have been related to a balance between excitatory and inhibitory inputs \cite{Amit01041997, vanVreeswijk1724} or to a disorder in the network or in the couplings 
\cite{Brunel2000, Cortes08102013} as crucial ingredients. %In networks of spiking neurons, 
Power law distributed avalanches have been attributed to synaptic 
plasticity with a stochastic noise in the charging \cite{Levina2007Dynamical, Bonachela2009, Bonachela2010, Millman2010a,volman2004generative} or to
dynamical mechanisms inspired by self organized criticality (SOC) \cite{PhysRevLett.96.028107,Chialvo2010,serena16}. 
The balance between excitation and inhibition plays an important 
role in the latter dynamical regime as well \cite{PhysRevLett.108.228703}, and 
a relation between uncorrelated dynamics in a network of stochastic units and  power law scaling has been proposed \cite{10.1371/journal.pone.0008982, Touboul2015}.

In this Letter we show that correlated irregular dynamics can be observed in homogeneous deterministic networks of $N$ identical purely excitatory spiking neurons endowed with synaptic plasticity, 
coupled by an all to all, mean field (MF), interaction. In this case, all neurons are synchronized but, for small enough synaptic decay time, the system displays 
a period doubling transition from a periodic phase to synchronous chaos \cite{ding,heagy}. 
Such a transition is determined by the competition among the system time scales in the strong and weak coupling limits.
For vanishing synaptic decay time, the dynamics can be reduced to a one dimensional map. 

In the presence of disorder in the couplings, we show that the dynamics exhibits three phases, depending on the interaction strength and synaptic decay time. In particular, next to the quasi-synchronous and the asynchronous regimes \cite{Burioni2014}, a  phase characterized by power law distributed avalanches emerges in correspondence to the chaotic phase  of the homogeneous MF model. Chaos is preserved in this dynamical phase, as confirmed by 
the computation of the Lyapunov exponents, and it is characterized by the onset of strong temporal correlations and high complexity. Our analysis uncovers a connection between dynamical stability and emergent avalanche activity in the presence of short-term synaptic plasticity, that may go beyond our particular case of study.

We consider a disordered random network of leaky integrate-and-fire (LIF) neurons \cite{Izhikevich2004} connected 
via the Tsodyks-Uziel-Markram (TUM) model for short term synaptic plasticity \cite{Tsodyks01012000}. Within a 
Degree based Mean Field approximation (DMF), for each neuron $i=1\dots N$  
the dynamics is defined by three differential equations: 
\begin{eqnarray}
  \dot{v}_{i}(t) &=& a - v_{i}(t) + g k_i Y(t)  \label{eq:vk}\\
  \dot{y}_{i}(t) &=& -\frac{y_{i}(t)}{\tau_{in}} + u ( 1 - y_{i}(t) - z_{i}(t) ) S_{i}(t)
  \label{eq:yk}
\end{eqnarray}

%\\
\begin{eqnarray}
\dot{z}_{i}(t) &=& \frac{y_{i}(t)}{\tau_{in}} - \frac{z_{i}(t)}{\tau_R}, \label{eq:zk}
\end{eqnarray}
where $v_{i}(t)$ is the membrane potential of neuron $i$ {while   
$y_{i}(t)$, $z_{i}(t)$ and $x_{i}(t)=1-y_{i}(t)-z_{i}(t)$ represent the active, inactive and available 
fraction of resources of the corresponding synapses. }
The potential $v_{i}(t)$ is reset to $0$ at times $t_{i}(m)$ when it reaches the threshold 
$v_{i}(t_{i}(m))=1$. At $t_{i}(m)$, a spike activates a fraction $u$ of the 
available resources, and the activation is modeled as a spike train 
 $S_{i}(t) = \sum_m \delta(t - t_{i}(m))$. 
Neurons are characterized by the coupling constant $g k_i$,  randomly extracted from the distribution $P(k_i)$. 
$N k_i$ can be interpreted as the effective number of neural synapses interacting with neuron $i$, 
i.e its in-degree \cite{Burioni2014}.
In this framework, $k_i$ is the only relevant topological feature of the 
neural network and it justifies the DMF name.
In a mean field description, the incoming synaptic current can be written as the average of the active resources
 $Y(t) = N^{-1} \sum_{i=1}^N y_{i}(t)$.
 
 {By introducing an event driven map \cite{Brette2007}, the DMF approach allows for very effective numerical simulations and it has been shown to reproduce
 	the relevant collective dynamics for networks with large finite connectivity and metrical features \cite{PhysRevE.90.022811} (see Supplemental Material (SM) \cite{supp}).}
 
Eqs. (\ref{eq:vk}-\ref{eq:zk})  are characterized by three time scales: 
the period of the oscillating non interacting neuron $T=\log(a/(a-1))$, the recovery time 
$\tau_R$ and the synaptic decay time $\tau_{in}$.  
 The regime  $\tau_{in} \lesssim T$ has been 
studied in detail in \cite{Burioni2014,PhysRevE.90.022811,PhysRevE.87.032801,PhysRevE.90.042918}, and it features a transition from a quasi-synchronous  to an 
asynchronous  phase as a function of $g$ and of the shape of $P(k_i)$.
Here we will focus instead on the regime $\tau_{in} \ll T \ll \tau_R$, setting $a=1.3$,  
$\tau_R=10$ and varying $\tau_{in}$ between $10^{-1}$ and $10^{-5}$. 
 These parameters are consistent with those selected in \cite{Tsodyks01012000}, where they have been
chosen on the basis of biological motivations.

{\it Mean Field.}  The presence of a further non trivial phase can be put into evidence by considering the simple 
in-degree distribution $P(k_i) = \delta(k_i - k_0)$.  In this fully MF case, where all the coupling constants are equal, all neurons become completely synchronized after an initial transient state, { as shown in the SM.}  Hence, Eqs. (\ref{eq:vk}-\ref{eq:zk}) reduce to the equations of a single neuron with coupling
$k_0$ and $Y(t)=y (t).$ The dynamics can be rewritten as an event 
driven Poincar\'e map in $z_n$ and $y_n$, representing the inactive and active resources before the $n$-th synchronous 
spiking event (see SM):
\begin{eqnarray}
y_{n+1} &=& e^{- \frac{\Delta_{n}}{\tau_{in}} } (y_n+u(1-y_n-z_n))
\label{eq:ykmap}
\\
z_{n+1} &=& -e^{- \frac{\Delta_{n}}{\tau_{in}} } \frac{ y_n+u(1-y_n-z_n) }{1-\tau_{in}/\tau_R} \nonumber\\
& & +e^{- \frac{\Delta_{n}}{\tau_{R}} }
\left( z_n + \frac{ y_n+u(1-y_n-z_n) }{1-\tau_{in}/\tau_R} \right), \label{eq:zkmap}
\end{eqnarray}
where the time interval $\Delta_n$ between  the $n$-th and the ($n+1$)-th spiking event is obtained
from: 
\begin{eqnarray}
1 &=& a - e^{- \frac{\Delta_{n}}{\tau_{in}} } \frac{ g \tau_{in} k_0(y_n+u(1-y_n-z_n)) }{1-\tau_{in}} \label{eq:vkmap} \\
& & - e^{-\Delta_{n}} \left( a - \frac{ g \tau_{in} k_0(y_n+u(1-y_n-z_n)) }{1-\tau_{in}} \right).   \nonumber 
\end{eqnarray}

When $\tau_{in}\ll T\ll \tau_R$, an insight on the dynamics 
can be achieved by considering the opposite regimes of weak and strong interaction, i.e. when 
$g k_0 Y(t)$ or $a-v_{ k_0}(t)$ are negligible in Eq. (\ref{eq:vk}), respectively.
In both extreme regimes, the map in Eqs. (\ref{eq:ykmap}-\ref{eq:vkmap}) can be solved, and it features
a fixed point corresponding to a periodic solution in the continuous dynamics 
(see SM for details). In particular, in the weak coupling regime, the periodicity 
is trivially $T$, and the interaction term remains negligible if $g  k_0 \tau_{in}\ll \tau_R /T$. 
On the other hand, if the $a-v_{ k_0}(t)$ term can be ignored, the system  
displays a much faster periodicity: $T_f=\tau_R/(g k_0 \tau_{in})$ and the approximations holds only if 
$g k_0 \tau_{in}\gg \tau_R /\tau_{in}$.

\begin{figure}
	\centering
	\includegraphics[scale=0.67]{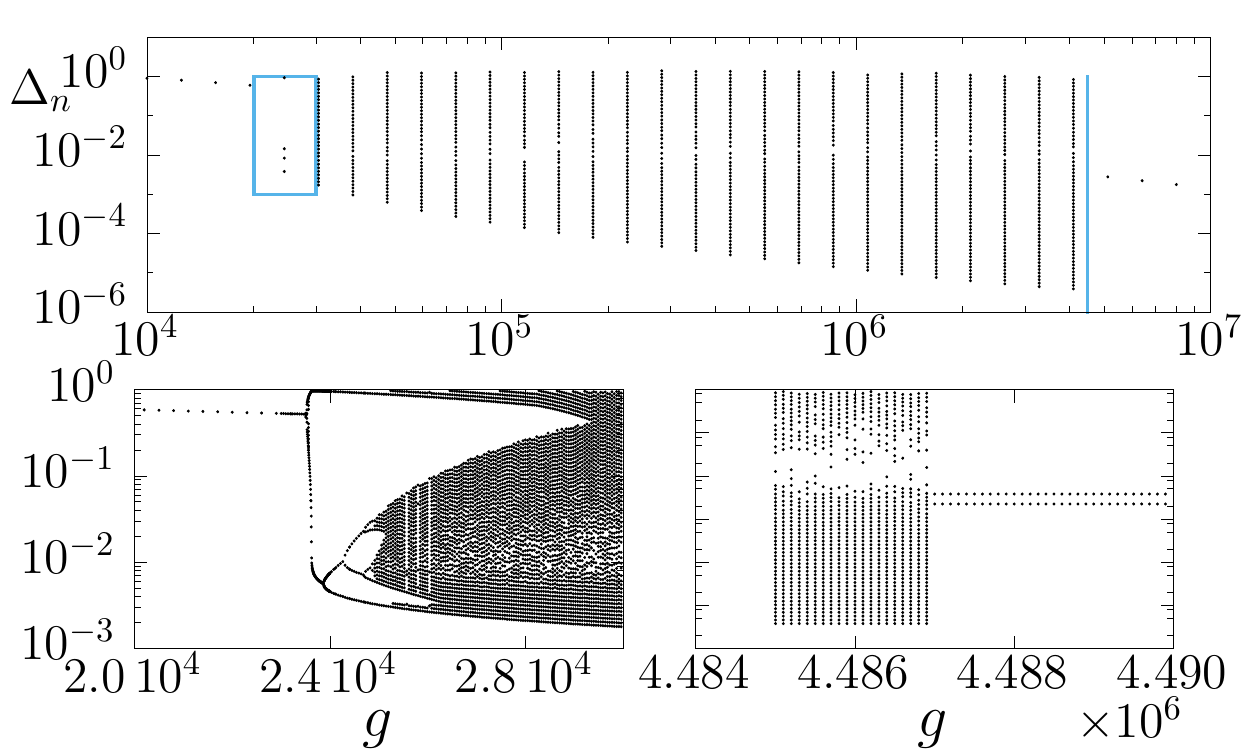}
	\caption{ \label{fig:pdMF}  Feigenbaum bifurcation diagram for the MF TUM model 
		in Eqs. (\ref{eq:ykmap}-\ref{eq:zkmap})
		with $\tau_{in} = 10^{-3}$. The attractor for the interspike interval of the network 
		$\Delta_n$ is showed as a function of %the chosen bifurcation parameter, i.e. 
		the coupling $g$. Upper panel: bifurcation diagram in the full relevant range of the 
		parameter $g$. Lower-left panel: magnification on the period doubling 
		cascade at the first transition.
		Lower-right panel: magnification on the second transition. The blue rectangles in the upper panel indicate the zooming regions of the lower panels.}
\end{figure}

\begin{figure}
	\centering
	\includegraphics[scale=0.63]{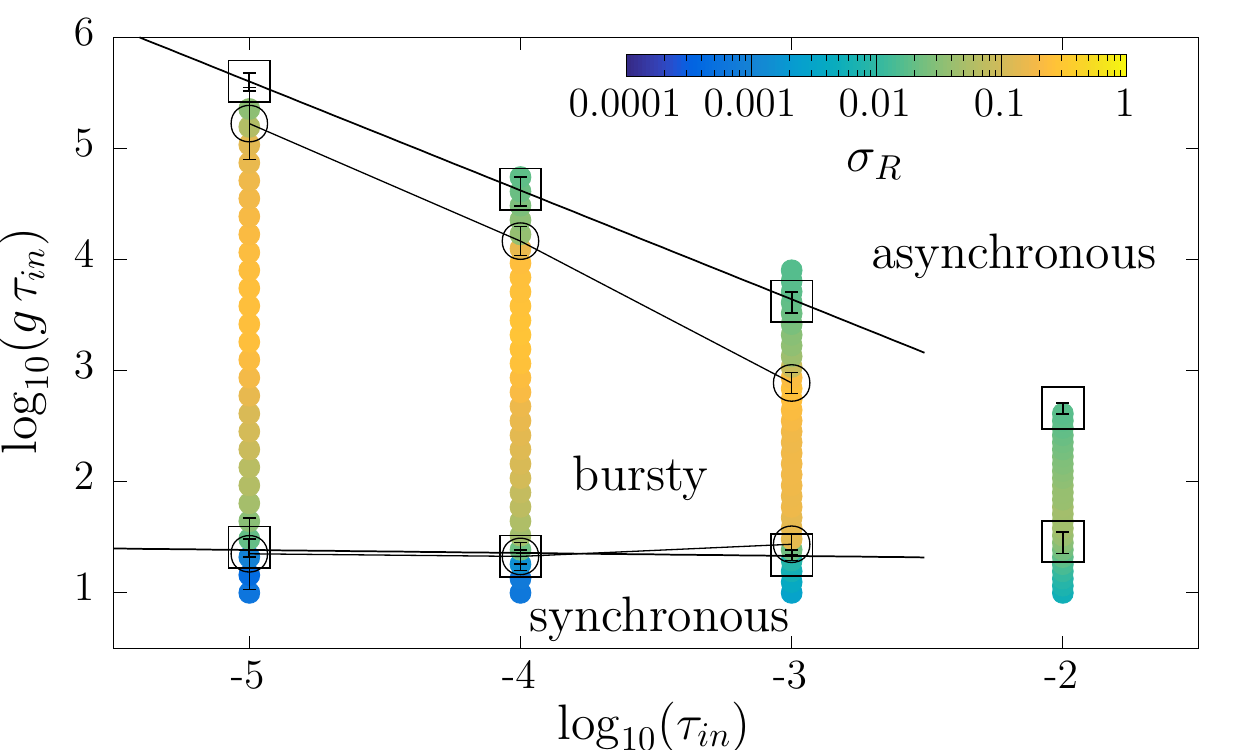}
	\caption{\label{fig:phaseD}(Color online) Dynamical phase diagram of the  MF and DMF 
		models in terms of the coupling constants 
		$g$ and of the synaptic time scale $\tau_{\rm in}$. 
		MF model: The squares indicate the $g$ values at which the transition to chaos (along with the discontinuity of the interspike time standard deviation) takes place (see SM). The black lines are linear fits. DMF model:  Each colored point corresponds to a simulation, the color code indicating $\sigma_R$ at the corresponding value of $(g,\tau_{\rm in})$. The intervals of $g$ containing the discontinuity (c.f. Fig. \ref{fig:K_HMF}) are signaled with black circles.}
\end{figure}

If $\tau_R /T \ll g k_0 \tau_{in} \ll  \tau_R /\tau_{in} $,
neither the weak nor the strong coupling conditions are satisfied, 
and the competition between the terms with a slow and a fast dynamics 
plays a non trivial role, destroying the presence of a periodic evolution.
Such a behavior can be analyzed by means of the bifurcation diagram \cite{ott} of $\Delta_n$ as a function of $g$ at fixed $\tau_{in}$. Fig. \ref{fig:pdMF} shows the presence of a stable 
fixed point for small and large values of $g$, describing a slow and a fast periodic regime, 
respectively. For an intermediate value, a period doubling appears first; 
then, at $g>g'(\tau_{in})$, the distribution of $\Delta_n$ becomes continuous.
The $\Delta_n$ becomes again delta-distributed for $g>g''(\tau_{in})$. 
{ In the SM we show that for $g'(\tau_{in})<g<g''(\tau_{in})$ the maximum Lyapunov exponent \cite{benettin1980lyapunov}
becomes positive, a signature of the presence of chaos. In the fully MF system with $N$ neurons, this is an example of synchronous chaos \cite{ding,heagy}.}  The phase diagram in Fig. \ref{fig:phaseD} shows that the $\tau_{in}$-dependence of the 
boundaries of the chaotic phase (squares) is consistent with the continuous lines, obtained by the weak and strong coupling limit arguments. {The critical values for $g$ and $\tau_{in}$ depend on $a$, i.e. the intrinsic period of the neuron; the chaotic dynamics is observed at higher $\tau_{in}$ by considering smaller $a$ (see SM).} 
Taking the limit $\tau_{in} \to 0$ with $g_{\rm eff}=g k_0 \tau_{in}$ constant in  
Eqs. (\ref{eq:ykmap}-\ref{eq:vkmap}), one obtains a single variable map as a 
function of $g_{\rm eff}$, $a$ and $\tau_R$ only, that can be studied analytically 
(see SM). This simpler map
confirms the presence of a genuine chaotic dynamical phase.

\begin{figure}
	\includegraphics[scale=0.64]{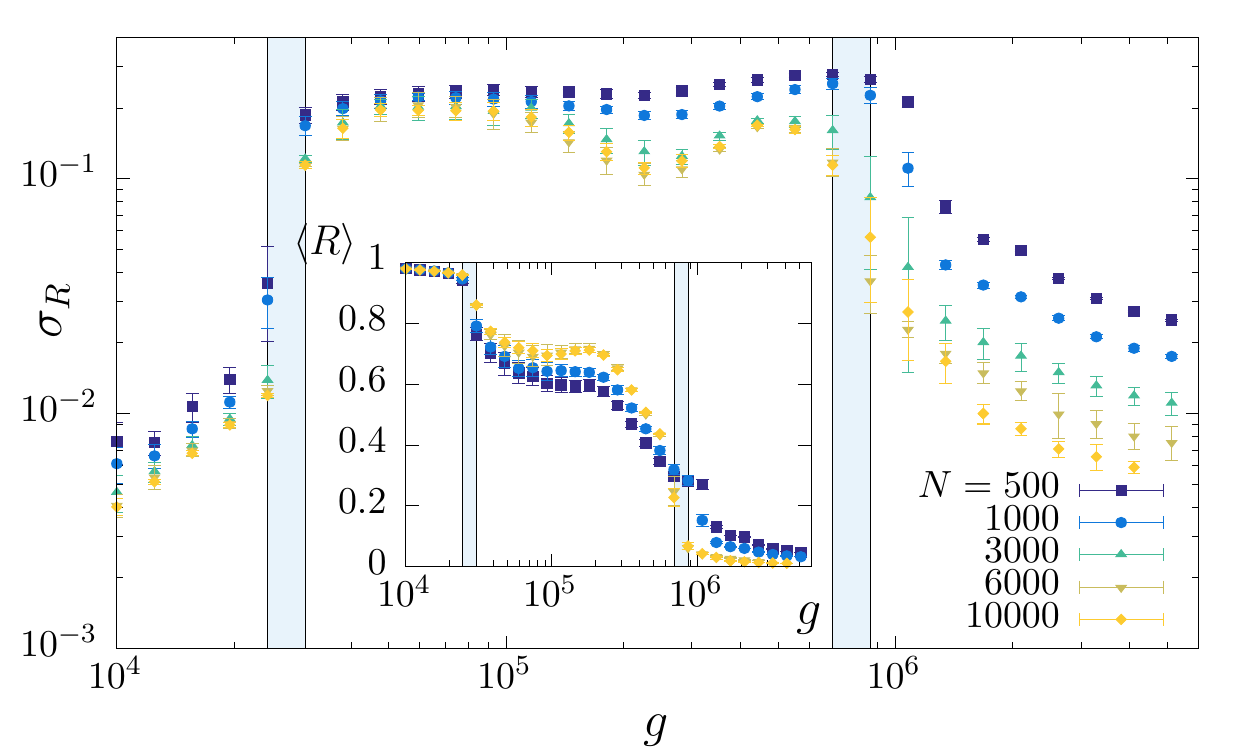}
	\centering
	\caption{\label{fig:K_HMF}(Color online) Standard deviation of the Kuramoto 
		parameter, $\sigma_R$ versus $g$ for the DMF model with $\tau_{in}=10^{-3}$ and five values of $N$. 
		In the quasi-synchronous and bursty  phases the data corresponding to the two larger values of $N$ 
		overlap within their statistical errors, indicating convergence in size, while 
		deep in the asynchronous phase they decrease as $\sim N^{-1/2}$. 
		Inset: temporal average of $R(t)$, showing that the larger sizes have attained 
		their asymptotic value in all the phases. The vertical stripes are common to 
		all the figures in the article and indicate the apparent discontinuity of 
		$\langle R \rangle$ for the largest sizes. }
\end{figure}

\begin{figure}
	\includegraphics[scale=0.64]{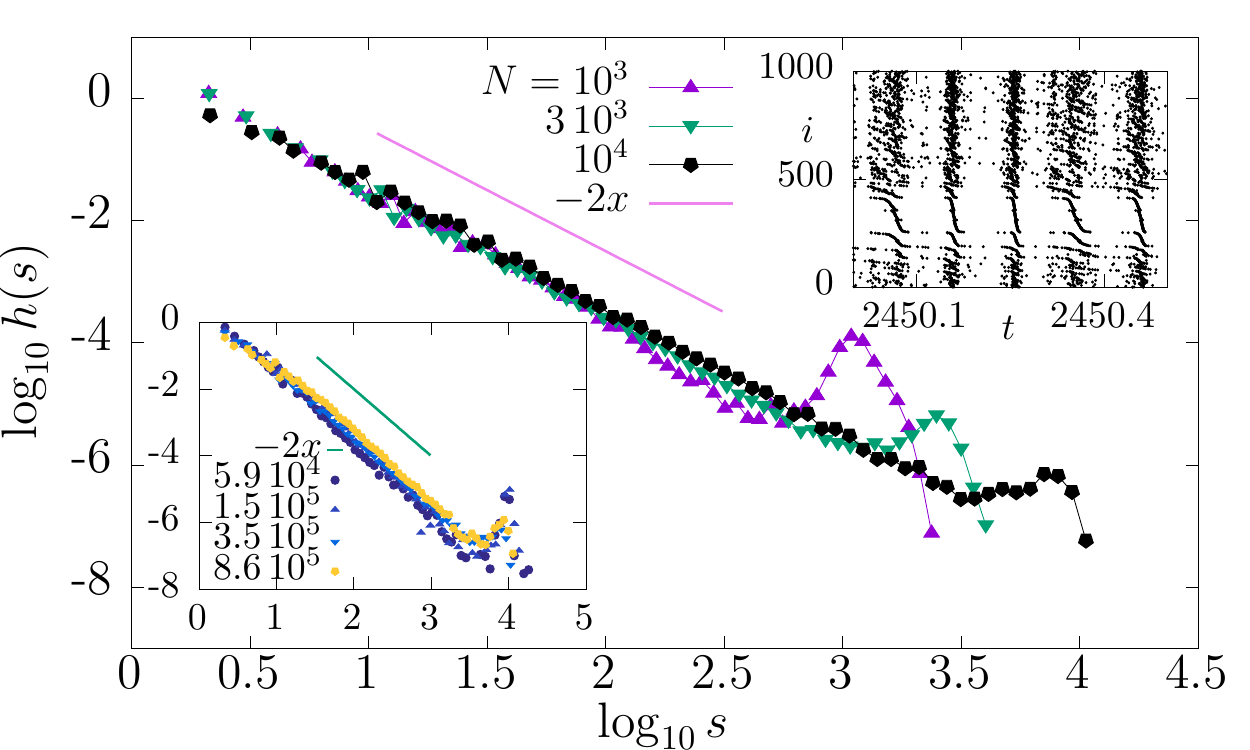}
	\centering
	\caption{\label{fig:avalanch}(Color online) Avalanche size histogram $h(s)$ of the DMF model, $\tau_{in}=10^{-3}$ and $g=3.5\cdot 10^{5}$ in the bursty regime, for several values of 
		$N$. % (see the definition of avalanche in the Supplemental Materials). 
		Upper inset: a fragment of the raster plot for the same system. 
		Lower inset: $\log_{10}h(s)$ for $N=10^4$ and various values of $g$ across the bursty phase.
	}
\end{figure}

{\it Degree based Mean Field.} Let us now focus on the  multi-site DMF model with heterogeneous couplings extracted from the distribution $P(k_i)$. {We consider a Gaussian $P(k_i)$ with average $\mu = 0.7$ and standard deviation $\sigma = 0.077$, although our results are robust for different distributions (see SM for a discussion).}
A relevant quantity describing 
the level of synchronization  of the neurons is the Kuramoto parameter \cite{RevModPhys.77.137}:
%\begin{eqnarray}
$
R\left(t\right)=\frac{1}{N}\left|\sum_{i=1}^{N}\textrm{e}^{\imath \phi_{i}\left(t\right)}\right|
$
%\end{eqnarray}
where $\phi_i(t)$ is the phase of neuron $i$ at time $t$:
\begin{eqnarray}
\phi_i(t) = 2 \pi \frac{t-t_{i}(m)}{t_{i}(m+1)-t_{i}(m)},
\label{kuraphase}
\end{eqnarray}
where $t_i(m)$ is the $m$-th spike of neuron $i$ and $t \in [t_i(m),\,t_i(m+1)]$. 
In Fig. \ref{fig:K_HMF} the time average $\langle R \rangle$ of the Kuramoto parameter 
and its fluctuations $\sigma_R$ are displayed as a function of $g$. At small couplings, 
$\langle R \rangle\approx 1$  and the fluctuations are small, as the systems is in a 
{\it quasi-synchronous} phase. At large $g$, $\langle R \rangle$ becomes very small 
($\langle R \rangle \rightarrow 0$ with increasing $N$), 
consistently with a periodic {\it asynchronous} phase. 
In the irregular, {\it bursty}, regime, $\langle R \rangle$ exhibits moderate values and, more significantly, its fluctuations grow abruptly by an order of magnitude; this is a signal of a complex dynamical phase, illustrated in the raster plot in  the inset of Fig. \ref{fig:avalanch} (each dot corresponds to a spike of neuron $i$ at time $t$). The fluctuations of $\langle R \rangle$ originate from 
 the alternations of synchronous events with asynchronous phases characterized 
 by smaller bursts where only a subset of the neurons fires simultaneously. 
The main plot of Fig. \ref{fig:avalanch} shows that the size $s$ of such bursts, or avalanches, is broadly distributed (see SM
 for a detailed definition
 of burst size). Interestingly, the distribution is compatible with a 
 power law $h(s)\sim s^{-\gamma}$ followed by a bump. The power $\gamma$, close to $2$ (see SM), does not depend significantly on $N$, nor on $g$ for a wide $g$-range in the bursty phase. Finally, the peaks at large 
 $s$ in the distributions correspond to synchronous events where all neurons 
 fire quasi-simultaneously, and their position scales with the system size.

The natural issue is the relation between the chaotic phase in the single site MF model and 
the bursty-avalanche regime of the multi-site DMF approach. In the SM we show that also the bursty phase is characterized by 
a chaotic dynamics with positive Lyapunov exponents. In Fig. \ref{fig:phaseD} we have superimposed the dynamical phase diagrams of the MF and DMF models. In the DMF, the transitions points (circles)   are set at the $g$ intervals at which the abrupt increments of the fluctuations of the Kuramoto parameter take place (c.f. Fig. \ref{fig:K_HMF}). In the MF case, the squares indicate the values of $g$ at which the transitions to chaos occur. While the phase diagrams slightly differ, the phase diagram of the DMF model converges continuously to that of the MF model in the limit of vanishing width of the distribution $P(k_i)$, as illustrated in the SM. {This scenario suggests that the bursty regime arises from the introduction of disorder on a system with synchronous chaos, so that neurons with different coupling  $k_i$ do not fire simultaneously and the synchronous solution loses stability}.

\begin{figure}
\includegraphics[scale=0.67]{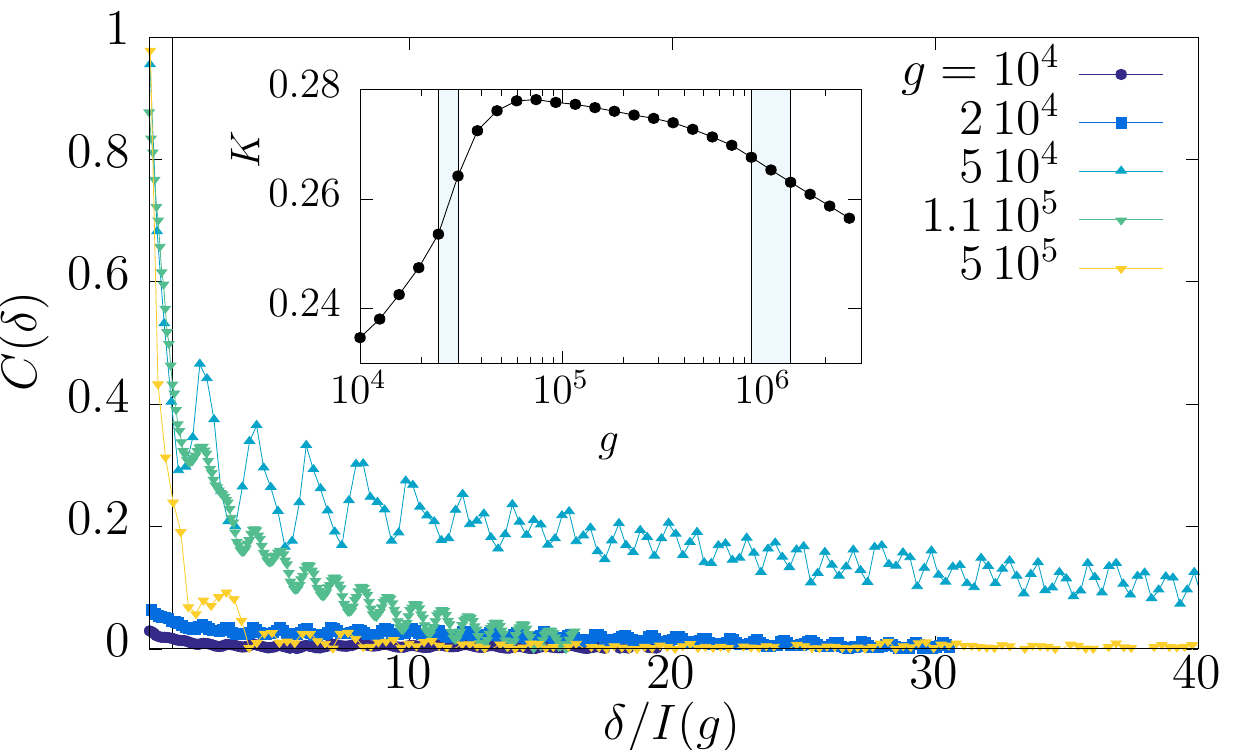}
\centering
\caption{\label{fig:Kcompl}(Color online) Main plot:  
connected correlation function $C$ as a function of the time 
difference $\delta$ in units of the average interspike time $I(g)$, for several 
values of $g$ in the DMF model with $\tau_{in}=10^{-3}$, $N=800$. 
For values of $g$ in the bursty phase, the correlation remains high even 
after large time differences.
Inset: 
Kolmogorov Complexity of the 
DMF model, $\tau_{in}=10^{-3}$, $N=10^3$, and the spike time differences $\Delta_n$ stored with 14 
digits of precision.
}
\end{figure}

In the DMF model, the transition to the bursty collective behavior also
corresponds to the presence of large temporal correlations.
 We define the time dependent {\it complex correlation}, 
$c(\delta, t)=(1/N)\sum_{i=1}^N{e^{\imath \phi_i(t)}e^{-\imath\phi_i(t+\delta)}}$, 
where $\phi_i(t)$ is the Kuramoto phase \eqref{kuraphase}, along with the 
{\it connected correlation function}, 
$C(\delta)=|\langle c(\delta,t)\rangle_t| -|\langle c({\cal T}, t) \rangle_t|$, 
as the temporal average $\langle\cdot\rangle_t$ of $c$ over a sufficiently large interval of times $t$, 
minus its stationary value at a sufficiently large time difference, $\delta={\cal T}$ 
(for details at this regard see the SM section). 
$C(\delta)$ measures in this way the average amount of correlation between  spike configurations separated by  a time delay $\delta$.
The quantity $C(\delta)$ (see Fig. \ref{fig:Kcompl}, main panel) reveals 
the existence of large correlations for times $\delta$ much larger than the 
average interspike time, $I(g)$, only in the bursty regime 
(for $3\cdot 10^4\lesssim g\lesssim 10^{5}$ at $\tau_{in}=10^{-3}$), 
while in the synchronous and asynchronous regimes, $C(\delta)$ decays faster to its 
asymptotic value.

Another interesting quantity in temporal series of 
neural firing patterns is the amount of information they can sustain.
In information theory, the Kolmogorov Complexity (KC) of a data sequence 
determines the length of the minimum computer program generating it, hence 
being a measure of the sequence predictability \cite{li2013}. KC has 
been related to the computational power of artificial neural networks 
\cite{Balcazar1997}, and used in the quantitative characterization of epileptic 
EEG recordings \cite{Petrosian1995}. We consider the KC of the raster plot, 
interpreting it as an estimation of the amount of information that can be codified 
in the dynamical signal (see the details of the KC estimation in the SM section). 
The numerical results for the DMF model reveal that the KC as a function of 
$g$ (see the inset of Fig. \ref{fig:Kcompl}) presents a maximum in the bursty regime (around $g\simeq 6\,10^4$ for ${\tau_{in}=10^{-3}}$). 

In summary, we have reported the existence of a dynamical phase occurring in 
a network of purely excitatory LIF neurons connected with synaptic plasticity. {This phase, identified by
	average statistical properties of the Kuramoto parameter, is 
	strongly chaotic and it differs from previously known irregular phases for similar models, e.g. 
	phases with chaotic transient dynamics \cite{Cortes08102013,zillmer2006desynchronization}.} 
The chaotic phase must also be distinguished from previous irregular regimes observed 
in spiking neural models, namely $\emph{weak chaos}$ in purely excitatory disordered networks 
\cite{PhysRevE.81.046119} or $\emph{stable\,chaos}$ in inhibitory ones \cite{PhysRevE.79.031909,ullner2016self, PhysRevLett.116.238101}.
The emergent dynamical regime occurs in a large region of the 
phase diagram, and it is separated by two dynamical transitions from the quasi-synchronous and asynchronous regimes. 
Chaos is preserved in the presence of disordered couplings. In that case, interestingly, the chaotic phase 
also features characteristic power law distributed avalanches. By properly defining temporal correlations and tools from information theory, we 
show that the additional bursty phase is strongly correlated and it carries a relevant amount 
of information compared to the quasi-synchronous and the asynchronous phases.

\begin{acknowledgments}
We gratefully acknowledge the support of
NVIDIA Corporation with the donation of the Tesla K40 GPU used for this research. We warmly thank S. di Santo, R. Livi, M. A. Mu\~noz and A. Politi for useful discussions.
\end{acknowledgments}

%\nocite{*}
\bibliographystyle{apsrev4-1}
%

%\bibliography{biblio3.bib}% Produces the bibliography via BibTeX.

%%%%% Including the supplemental material

%\foreach \x in {1,...,26}
%{%
%\clearpage
%\includepdf[pages={\x,{}}]{supplemental_arxiv_v2.pdf}
%}

\clearpage
\newpage
\thispagestyle{empty}
\setcounter{page}{1}
\foreach \x in {1,...,26}
{%
  \clearpage
  \thispagestyle{empty}
  \centering
  \includegraphics[page=\x, scale=0.9]{supplemental_arxiv_v2.pdf}
}

\end{document}